# Dynamic scaling properties of multistep polarization response in ferroelectrics

Y. A. Genenko[1*], S. Zhukov[2], M.-H. Zhang[3], K. Wang[4], and J. Koruza[5]

[1]*Materials Modelling, Department of Materials and Earth Sciences, Technical University of Darmstadt, Otto-Berndt- Str. 3, 64287 Darmstadt, Germany*

[2]*Department of Materials and Earth Sciences, Technical University of Darmstadt, Otto-Berndt-Str. 3, 64287 Darmstadt, Germany*

[3]*Nonmetallic Inorganic Materials, Department of Materials and Earth Sciences, Technical University of Darmstadt, Alarich-Weiss-Straße 2, 64287 Darmstadt, Germany*

[4]*State Key Laboratory of New Ceramics and Fine Processing, School of Materials Science and Engineering, Tsinghua University, Beijing, China*

[5]*Institute for Chemistry and Technology of Materials, Graz University of Technology, Stremayrgasse 9, 8010 Graz, Austria*

## Abstract

Ferroelectrics are multifunctional smart materials finding applications in sensor technology, micromechanical actuation, digital information storage etc. Their most fundamental property is the ability of polarization switching under applied electric field. In particular, understanding of switching kinetics is essential for digital information storage. In this regard, scaling properties of the temporal polarization response are well-known for 180°-switching processes in ferroelectrics characterized by a unique field-dependent local switching time. Unexpectedly, these properties were now observed in multiaxial polycrystalline ferroelectrics, exhibiting a number of parallel and sequential non-180°-switching processes with distinct switching times. This behaviour can be explained by a combination of the multistep stochastic mechanism and the inhomogeneous field mechanism models of polarization reversal. Scaling properties are predicted for polycrystalline ferroelectrics of tetragonal, rhombohedral and orthorhombic symmetries and exemplarily demonstrated by measurements of polarization kinetics in (K,Na)NbO$_3$-based ferroelectric ceramic over a timescale of 7 orders of magnitude. Dynamic scaling properties allow insight into the microscopic switching mechanisms, on the one hand, and into statistical material characteristics, on the other hand, providing thereby the description of temporal polarization with high accuracy. The gained deeper insight into the mechanisms of multistep polarization switching is crucial for future ultrafast and multilevel digital information storage.

---

* Corresponding author: genenko@mm.tu-darmstadt.de

## 1. Introduction

The most characteristic feature of ferroelectric materials is the appearance of spontaneous polarization below the ferroelectric-paraelectric phase transition temperature. Spontaneous polarization can be switched to another direction by applying an electric field or mechanical stress. The temporal characteristics of the response to an applied electric field are crucial for various applications of these materials, in particular ferroelectric memories (FeRAM) [1], future multi-level data storage [2-5] and ferroelectric field-effect transistors (FeFET) assembled in neuromorphic computing systems [6]. The field-driven polarization kinetics in single-crystalline ferroelectrics was successfully described by the stochastic Kolmogorov-Avrami-Ishibashi (KAI) model [7-9] assuming the formation of 180°-reversed polarization domains characterized by a unique field-dependent switching time. This model, however, failed when applied to thin ferroelectric films and bulk ferroelectric ceramics, both of which exhibit a broad statistical distribution of switching times [10-13]. To account for the dispersive character of the polarization response, the Nucleation Limited Switching (NLS) model was developed by Tagantsev *et al*. [10] and subsequently applied successfully to lead zirconate titanate (PZT) thin films [11]. The NLS model, however, did not reveal the physical origin of the wide distribution of switching times. To explain this feature, an idea was suggested by Lupascu *et al*. [12] that the statistical distribution of switching times may result from the random spatial distribution of an applied field in polycrystalline ferroelectrics, considering the well-known strong field dependence of the switching time [14]. This idea was first implemented quantitatively by Jo *et al*. using the Lorentzian statistical distribution of local field values in the NLS approach and successfully applied to PZT thin films [15,16] and organic ferroelectrics [17-19] and composites [20,21]. Alternatively, the Gaussian field distribution was proposed for bulk PZT ceramics [22], but also seems to be suitable for HfO$_2$-based thin films [6,23-28]. It was established, however, that there is no need to *a priori* assume a statistical distributions of local fields because they can be extracted directly from the polarization response, as will be explained below.

Many organic and inorganic polycrystalline ferroelectrics were found to possess scaling properties of their temporal polarization response to an applied electric field, combining dependences on the field and time variables. These materials include PZT [29] and $(Ba_{0.85}Ca_{0.15})TiO_3$ (BCT) [30] ceramics of different phase symmetries, $0.94(Bi_{1/2}Na_{1/2})TiO_3$–$0.06BaTiO_3$ (BNT-BT) [31] and $(1-x)Ba(Zr_{0.2}Ti_{0.8})O_3$-$x(Ba_{0.7}Ca_{0.3})TiO_3$ (BZT-BCT) [32] solid solutions of different phase symmetries, organic ferroelectric copolymers of vinylidenefluoride and trifluoroethylene P(VDF-TrFE) [33], and $HfO_2$-based thin films [23-28]. The scaling property is expressed by the fact that the normalized derivative of the polarization variation $\Delta P(E_a, t)$ with respect to the applied field, $E_a$, can be represented as a function of combined variable of $E_a$ and time $t$,

$$\frac{E_a}{\Delta P_{max}} \frac{\partial \Delta P(E_a, t)}{\partial E_a} = \Phi\left(\frac{E_a}{E_{max}(t)}\right) \qquad (1)$$

where the maximum polarization variation, $\Delta P_{max}$, the maximum position of the polarization derivative with respect to the field, $E_{max}(t)$, and the shape of a master curve, $\Phi(s)$, are unique characteristics of a certain material. The relation (1) means that all normalized derivatives (1) evaluated for different times $t$ fall on the same master curve when the field argument is normalized to $E_{max}(t)$. If the relation (1) is experimentally confirmed, that is the case for all the materials mentioned above [23-33], then the polarization response can be described with high accuracy by a formula [34]

$$\Delta P(E_a, t) = \Delta P_{max} \int_0^{E_a/E_{max}(t)} \frac{du}{u} \Phi(u) \qquad (2)$$

which applies over a wide time domain for any field value $E_a$.

Explanation of this behavior was given by the Inhomogeneous Field Mechanism (IFM) model [34,35] based on the following assumptions:

- The field-driven polarization reversal proceeds via statistically independent one-step 180°-switching events (as assumed in the KAI model)
- During the polarization reversal a stable statistical distribution of local fields $F(E/E_a, \theta)$ remains, which depends on the polar angle $\theta$ with respect to the applied field direction and the local field amplitude, $E$, normalized to the applied field $E_a$

- Local polarization switching is characterized by a switching time $\tau(E)$ dependent on the local electric field magnitude $E$
- The KAI time dependence for local polarization switching, $1 - \exp\left[-(t/\tau)^\beta\right]$, with the Avrami index $\beta$ [8,9], is approximated by the Heaviside step-function $\vartheta(t - \tau)$.

The assumption of the stable statistical field distribution $F(E/E_a, \theta)$ means neglecting the emerging depolarization fields, a hypothesis supported by the recent self-consistent simulations of polarization reversal in polycrystalline ferroelectrics [36]. The step-function approximation on the time scale is justified when the local polarization switching step is sharper than the statistical distribution of local fields [34], therefore it works well in polycrystalline systems, including highly textured ceramics [30], but fails in single crystals [37].

Having the master curve $\Phi(u)$ established, a weighted statistical distribution function for field values,

$$f(s) = \frac{1}{\langle \cos\theta \rangle} \int_0^\pi d\theta \sin\theta \cos\theta F(s, \theta), \tag{3}$$

can be derived as [34]

$$f(s) = \frac{1}{s} \Phi\left(\frac{1}{\gamma s}\right) \tag{4}$$

with $\gamma$ a constant explained later and the mean cosine of $\theta$ defined by the relation

$$\Delta P_{\max} = 2P_s \langle \cos\theta \rangle = 2P_s \int_0^\infty ds \int_0^\pi d\theta \sin\theta \cos\theta F(s, \theta). \tag{5}$$

A conceptual problem appears when considering polarization reversal in multiaxial ferroelectrics, as was noted in Ref. [38]. As was established earlier by X-ray diffraction [39] and ultrasonic measurements [40] on tetragonal ferroelectric ceramics, polarization switching occurs at least partially by sequential 90°-reorientations. Distinct characteristic times of two consecutive non-180° switching steps were resolved by *in situ* X-ray diffraction studies [41,42]. The recently elaborated original Multistep Stochastic Mechanism (MSM) model allowed the identification of fractions, activation fields and switching times of sequential 90°-90°-switching processes in tetragonal ceramics [43], 109°-71°-, 71°-109°- and 71°-71°-71°-switching processes in rhombohedral ceramics [44], and 120°-60°-, 60°-120°-, 90°-90°- and 60°-60°-60°-switching processes in orthorhombic ceramics [45]. Considering a large number of polarization switching paths in multiaxial ferroelectrics with switching times of distinct switching events different by orders of the magnitude, the scaling property of the polarization response was not expected in these systems. Paradoxically, the scaling properties [29] and the multistep switching processes [46] were clearly observed in the same tetragonal and rhombohedral PZT compositions. In order to resolve this paradox, in the current study, the compatibility of the IFM and the MSM models is investigated. It is shown that, despite multiple parallel and sequential switching processes with different switching times, the scaling property can be preserved under certain conditions. This is proved theoretically for ferroelectrics of tetragonal, rhombohedral and orthorhombic phase symmetries and experimentally confirmed by the exemplary case of the polarization response of an orthorhombic $(K,Na)NbO_3$-based ceramic [45]. The theoretical analysis is presented in detail for the latter, most complicated case, while the derivations for tetragonal and rhombohedral cases are presented in a concise form in Appendices A and B, respectively.

## 2. Scaling theory of the multistep polarization reversal in orthorhombic polycrystalline ferroelectrics

In orthorhombic ferroelectrics with a high-temperature cubic parent phase, local spontaneous polarizations may be present in one of the twelve possible directions along the medial plane diagonals of the pseudo-cubic cell (<011> directions). A sufficiently strong electric field opposite to the initial polarization direction causes a reversal of polarization. Let us choose a Cartesian frame with axes collinear with main axes of a pseudo-cubic cell (Figure 1(a)). The initial polarization state is assumed to be $P_s(0, -1, -1)/\sqrt{2}$ (polarization of the amplitude $P_s$ pointing to A).

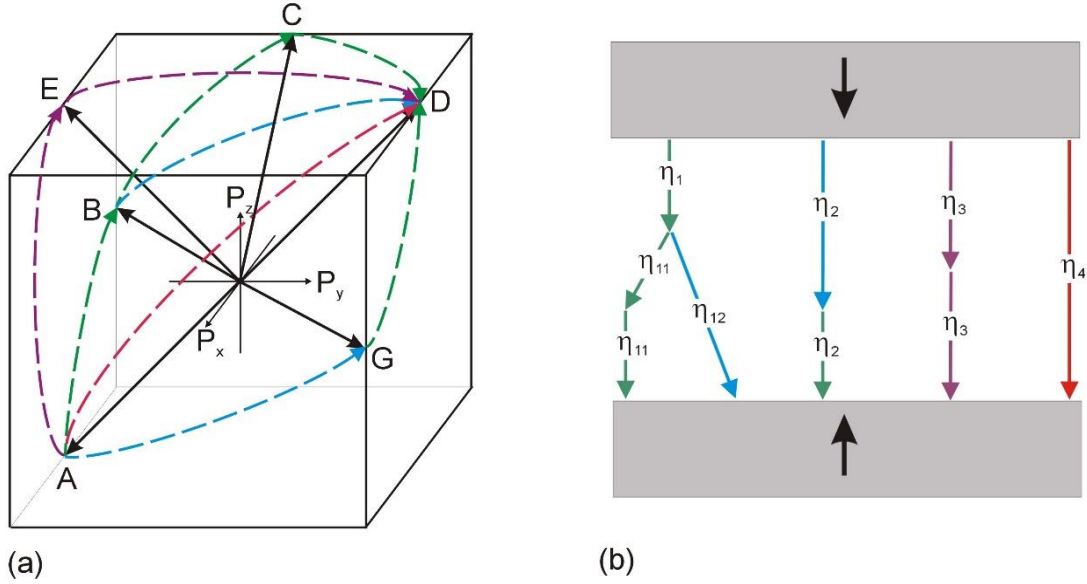

Figure 1. (a) Possible paths of the field-induced polarization reversal in orthorhombic state: exemplary 60°-60°-60° polarization rotation path A-B-C-D, exemplary 60°-120° polarization rotation path A-B-D, exemplary 120°-60° polarization rotation path A-G-D, exemplary 90°-90° polarization rotation path A-E-D, and a direct 180° polarization reversal A-D. (b) Definition of fractions of different switching processes. Both in (a) and (b) 60°-switching processes are indicated with green lines, 120°-switching processes with blue lines, 90°-switching processes with purple lines and 180°-switching processes with red lines. Arrows indicate the end of each process.

An electric field with opposite direction drives the spontaneous polarization to the final state $P_S(0,1,1)/\sqrt{2}$ (polarization pointing to D). Polarization reversal may occur along different paths exemplarily depicted in Figure 1(a), namely, by a direct 180° path A-D, by two sequential 120°- and 60°- polarization rotations A-G-D, by two sequential 60°- and 120°-polarization rotations A-B-D and by a triple sequential 60°- polarization rotation A-B-C-D. We indicate the probabilities for the first switching steps as $\eta_1$, $\eta_2$, $\eta_3$ and $\eta_4$ for processes starting with 60°, 120°, 90° and 180° rotations, respectively, i.e. $\eta_1+\eta_2+\eta_3+\eta_4=1$. Following the first 60°-polarization rotation, the two following paths are possible, a single 120°-polarization rotation with a weight $\eta_{12}$ (such as B-D path) and a double 60°-polarization rotation with a weight $\eta_{11}$ (such as B-C-D path), which satisfy $\eta_{11}+\eta_{12} = \eta_1$. In a stochastic approach, the parameters $\eta_{11}$, $\eta_{12}$, $\eta_2$, $\eta_3$ and $\eta_4$ are understood as average characteristics of the whole system.

Keeping the simplified KAI assumption that switching processes at different places are statistically independent, the probabilities of distinct switching steps in orthorhombic ferroelectrics were recently derived in the MSM approach [45]. At the beginning, we assume a single-domain single crystal with saturation polarization along $[0\bar{1}\bar{1}]$ (as in Figure 1(a)). When an opposite electric field is applied, local polarizations may undergo different switching processes with distinct nucleation rates per unit time and unit volume and develop growing switched domains of different dimensionalities. This results in the time dependent total polarization variation in the field direction,

$$\Delta P(t) = \Delta P_{180}(t) + \Delta P_{90-90}(t) + \Delta P_{120-60}(t) + \Delta P_{60-120}(t) + \Delta P_{60-60-60}(t). \tag{6}$$

In the following we assume, as in the IFM approach to polycrystalline ferroelectrics [35], the step-function approximation for the local polarization time dependences, and apply this to switching probabilities of particular switching paths derived in [45]. Thus, for the contribution of 180°-switching events,

$$\Delta P_{180}(t) = \eta_4 2P_s \left\{ 1 - \exp\left[ -\left(\frac{t}{\tau_4}\right)^{\beta_4} \right] \right\} \cong \eta_4 2P_s \vartheta(t - \tau_4). \tag{7}$$

with the respective Avrami index $\beta_4$ (related to the growing domain dimensionality) and switching time $\tau_4$ of this process. For the contribution of two-step 90°-90°-switching events,

$$\Delta P_{90-90}(t) = \eta_3 P_s \left[ L_{3\bar{3}}(t) + 2L_{33}(t) \right], \tag{8}$$

the switching probabilities to switch once on this path, $L_{3\bar{3}}(t)$, and two times, $L_{33}(t)$, by time $t$ are accordingly approximated as

$$L_{3\bar{3}}(t) = \frac{\beta_3}{\tau_3} \int_0^t dt_1 \left(\frac{t_1}{\tau_3}\right)^{\beta_3-1} \exp\left[-\left(\frac{t_1}{\tau_3}\right)^{\beta_3}\right] \exp\left[-\left(\frac{t-t_1}{\tau_{33}}\right)^{\beta_{33}}\right] \cong \vartheta(t-\tau_3) - \vartheta(t-\tau_3-\tau_{33}) \tag{9a}$$

$$L_{33}(t) = 1 - \exp\left[-\left(\frac{t}{\tau_3}\right)^{\beta_3}\right] - L_{3\bar{3}}(t) \cong \vartheta(t-\tau_3-\tau_{33}) \tag{9b}$$

with respective Avrami indices $\beta_3$, $\beta_{33}$ and switching times $\tau_3$, $\tau_{33}$ of the first and the second switching steps. This results in the total contribution of this switching path,

$$\Delta P_{90-90}(t) \cong \eta_3 P_s \left[ \vartheta(t-\tau_3) + \vartheta(t-\tau_3-\tau_{33}) \right]. \tag{10}$$

Completely analogously, for the other two-step switching paths,

$$\Delta P_{120-60}(t) = \eta_2 P_s \left[ \frac{3}{2} L_{2\bar{1}}(t) + 2L_{21}(t) \right] \cong \eta_2 P_s \left[ \frac{3}{2} \vartheta(t - \tau_2) + \frac{1}{2} \vartheta(t - \tau_2 - \tau_{21}) \right], \tag{11}$$

$$\Delta P_{60-120}(t) = \eta_{12} P_s \left[ \frac{1}{2} L_{1\bar{2}}(t) + 2L_{12}(t) \right] \cong \eta_{12} P_s \left[ \frac{1}{2} \vartheta(t - \tau_1) + \frac{3}{2} \vartheta(t - \tau_1 - \tau_{12}) \right]. \tag{12}$$

For the triple-step switching path one finds

$$\Delta P_{60-60-60}(t) = \eta_{11} P_s \left[ \frac{1}{2} L_{1\bar{1}}(t) + \frac{3}{2} L_{11\bar{1}}(t) + 2L_{111}(t) \right] \tag{13}$$

with approximations for probabilities for the first switching step, $L_{1\bar{1}}(t)$, the second switching step, $L_{11\bar{1}}(t)$, and the third switching step, $L_{111}(t)$,

$$L_{1\bar{1}}(t) \cong \vartheta(t - \tau_1) - \vartheta(t - \tau_1 - \tau_{11}),$$
$$L_{11\bar{1}}(t) \cong \vartheta(t - \tau_1 - \tau_{11}) - \vartheta(t - \tau_1 - \tau_{11} - \tau_{111}),$$
$$L_{111}(t) \cong \vartheta(t - \tau_1 - \tau_{11} - \tau_{111}),$$

resulting in approximated Eq. (13),

$$\Delta P_{60-60-60}(t) \cong \eta_{11} P_s \left[ \frac{1}{2} \vartheta(t - \tau_1) + \vartheta(t - \tau_1 - \tau_{11}) + \frac{1}{2} \vartheta(t - \tau_1 - \tau_{11} - \tau_{111}) \right]. \tag{14}$$

Using the formulas (10-12,14) in the following, we will account for the fact that switching times of the sequential switching processes differ from each other by orders of the magnitude [45], $\tau_{33} \ll \tau_3$, $\tau_{21} \ll \tau_2$, $\tau_1 \ll \tau_{12}$, $\tau_{111} \ll \tau_{11} \ll \tau_1$, so that the shorter times in additive expressions may be neglected.

When considering polycrystalline ferroelectrics in the spirit of the IFM model [34,35], we average the total polarization (6) over statistical distributions of local switching times, which are assumed to follow from the statistical distribution of local electric fields. Thereby the change from the time to the field scale is required. To this end, the formulas (7,10-12,14) are transformed to

$$\Delta P_{180}(t) \cong \eta_4 2P_s \vartheta(t-\tau_4) = \eta_4 2P_s \vartheta(E - E_{th}^{(4)}(t))$$
$$\Delta P_{90-90}(t) \cong \eta_3 2P_s \vartheta(t-\tau_3) = \eta_3 2P_s \vartheta(E - E_{th}^{(3)}(t))$$
$$\Delta P_{120-60}(t) \cong \eta_2 2P_s \vartheta(t-\tau_2) = \eta_2 2P_s \vartheta(E - E_{th}^{(2)}(t))$$
$$\Delta P_{60-120}(t) \cong \eta_{12} P_s \left[ \frac{1}{2}\vartheta(t-\tau_1) + \frac{3}{2}\vartheta(t-\tau_{12}) \right] \quad (15)$$
$$= \eta_{12} P_s \left[ \frac{1}{2}\vartheta(E - E_{th}^{(1)}(t)) + \frac{3}{2}\vartheta(E - E_{th}^{(12)}(t)) \right]$$
$$\Delta P_{60-60-60}(t) \cong \eta_{11} 2P_s \vartheta(t-\tau_1) = \eta_{11} 2P_s \vartheta(E - E_{th}^{(1)}(t))$$

where the introduced threshold fields $E_{th}^{(n)}(t)$ are the solutions of respective equations $\tau_n(E) = t$.

Now, by averaging over the statistical field distribution, $F(E/E_a, \theta)$, the total polarization variation results as

$$\frac{\Delta P(E_a,t)}{\Delta P_{\max}} \cong \eta_4 \int_{E_{th}^{(4)}/E_a}^{\infty} ds\, f(s) + \eta_3 \int_{E_{th}^{(3)}/E_a}^{\infty} ds\, f(s) + \eta_2 \int_{E_{th}^{(2)}/E_a}^{\infty} ds\, f(s)$$
$$+ \frac{3}{4}\eta_{12} \int_{E_{th}^{(12)}/E_a}^{\infty} ds\, f(s) + \left(\eta_{11} + \frac{1}{4}\eta_{12}\right) \int_{E_{th}^{(1)}/E_a}^{\infty} ds\, f(s) \quad (16)$$

with the weighted statistical field distribution, Eq. (3). One can get rid of the integral representation in Eq. (16) by differentiation with respect to the applied field $E_a$ resulting in the form

$$\frac{E_a}{\Delta P_{\max}} \frac{\partial \Delta P(E_a,t)}{\partial E_a} \cong \eta_4 \frac{E_{th}^{(4)}}{E_a} f\left(\frac{E_{th}^{(4)}}{E_a}\right) + \eta_3 \frac{E_{th}^{(3)}}{E_a} f\left(\frac{E_{th}^{(3)}}{E_a}\right)$$
$$\eta_2 \frac{E_{th}^{(2)}}{E_a} f\left(\frac{E_{th}^{(2)}}{E_a}\right) + \frac{3}{4}\eta_{12} \frac{E_{th}^{(12)}}{E_a} f\left(\frac{E_{th}^{(12)}}{E_a}\right) + \left(\eta_{11} + \frac{1}{4}\eta_{12}\right)\frac{E_{th}^{(1)}}{E_a} f\left(\frac{E_{th}^{(1)}}{E_a}\right). \quad (17)$$

The latter formula resembles the scaling form (1) and may also exhibit scaling properties under some prerequisites. Let us assume now that all polarization switching events proceed accordingly to the same physical mechanism and are characterized by similarly field-dependent switching times with distinct activation fields. For example, if they all exhibit similar Merz-like electric field dependences [14], the threshold fields obey the same time dependence with distinct factors (activation fields),

$$\tau_n(E) = \tau_0 \exp\left[\left(\frac{E_A^{(n)}}{E}\right)^\alpha\right] \Rightarrow E_{th}^{(n)} = \frac{E_A^{(n)}}{\left[\ln(t/\tau_0)\right]^{1/\alpha}}. \tag{18}$$

Thus, all the terms on the right-hand side of Eq. (17) are dependent on the same combined variable of the applied field and time, $E_a \left[\ln(t/\tau_0)\right]^{1/\alpha}$. Then the left-hand side of Eq. (17) must also be a function of the same combined variable. The dependence of the polarization derivative on the applied field typically reveals a peak at some field value $E_{\max}(t)$ which then has to follow the same time dependence. This provides the relations $E_{\max} = \gamma_1 E_{th}^{(1)} = \gamma_2 E_{th}^{(2)} = \gamma_3 E_{th}^{(3)} = \gamma_{12} E_{th}^{(12)} = \gamma_4 E_{th}^{(4)}$ with some constant factors $\gamma_n$. Finally, the derivative (17) may be represented in the scaling form (1) with the master curve

$$\begin{aligned}\Phi(s) &= \eta_4 \frac{1}{\gamma_4 s} f\left(\frac{1}{\gamma_4 s}\right) + \eta_3 \frac{1}{\gamma_3 s} f\left(\frac{1}{\gamma_3 s}\right) + \eta_2 \frac{1}{\gamma_2 s} f\left(\frac{1}{\gamma_2 s}\right) \\ &+ \frac{3}{4} \eta_{12} \frac{1}{\gamma_{12} s} f\left(\frac{1}{\gamma_{12} s}\right) + \left(\eta_{11} + \frac{1}{4}\eta_{12}\right) \frac{1}{\gamma_1 s} f\left(\frac{1}{\gamma_1 s}\right)\end{aligned} \tag{19}$$

followed by the expression for the time dependent polarization (2). We emphasize that, for using Eq. (2), the knowledge of process fractions $\eta_n$ and respective constants $\gamma_n$ is not required, as soon as the relation (1) is experimentally established. However, the process fractions have a great effect on strain response and can be established from analysis of dynamic strain data [45].

The scaling properties (1-2) are not expected to apply to each material and are subjected to a range of the mentioned assumptions concerning the width of the statistical distribution of local fields and the field dependence of switching times for different processes. However, they are expected to apply to a wide class of polycrystalline ferroelectrics that should be checked experimentally for each particular material. Indeed, similarly to the above considered ferroelectrics of orthorhombic symmetry, the materials of tetragonal and rhombohedral symmetries also exhibit dynamic scaling properties, though with different master curves $\Phi(s)$, which are derived in Appendices A and B, respectively. In the following, the scaling properties of an exemplary KNN-based ceramic [45] will be investigated experimentally.

## 3. Experimental proof of scaling properties of multistep switching

Bulk ceramics with nominal chemical composition $(Na_{0.49}K_{0.49}Li_{0.02})(Nb_{0.8}Ta_{0.2})O_3$, modified with 2 wt% $MnO_2$, were processed as described in [45] using the conventional solid-state reaction route [47]. X-ray powder diffraction revealing orthorhombic structure and standard electromechanical characterization of the sample combining large-signal polarization and strain measurements were presented in Ref. [45]. Time-dependent polarization reversal was measured using the pulse switching setup and methodology described in Ref. [42]. A broad range of electric field pulses, from $E_a = 0.15$ kV/mm to 1.30 kV/mm, which are approximately in the range of $0.3E_C < E_a < 2.6E_C$, were applied to the sample, where $E_C$ is the coercive field. Experimental data on time dependent polarization switching are presented by symbols in **Figure 2(a)** for exemplary field values. To study the scaling properties, the polarization variation data are first represented as a function of the applied field at different fixed poling time values in Figure 2(b). All curves exhibit inflection points related to the maxima of the first polarization derivatives with respect to the field. These derivatives are presented in Figure 2(c) using again the time as a parameter. From Figure 2(c), a position of maximum, $E_{max}(t)$, can be derived which is shown in Figure 3. By calculating the normalized derivatives (1) for different times, using the experimental value $\Delta P_{max}$ = 36.1 µC/cm², and their representation vs. the applied field normalized to $E_{max}(t)$, a master curve $\Phi(u)$ for the studied KNN-ceramic is established in Figure 2(d) thus confirming the scaling property of polarization response in this material.

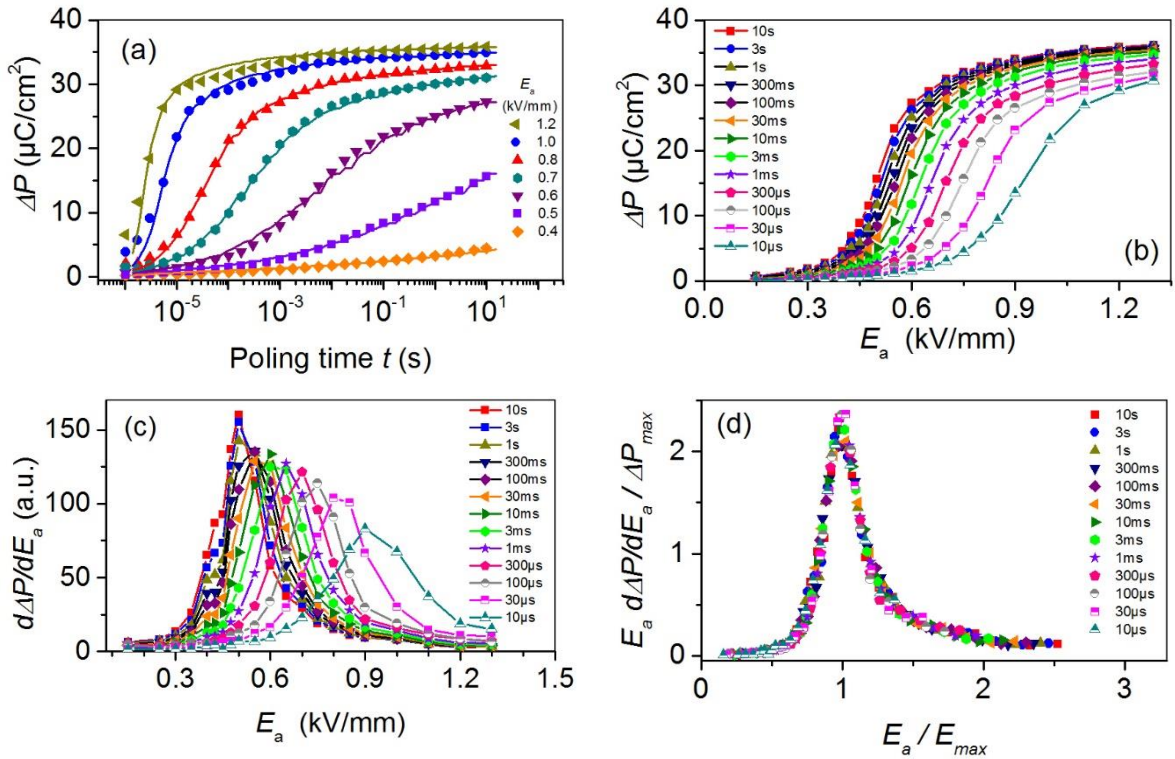

Figure 2. (a) Variation of polarization with time during field-induced polarization reversal at different applied field values in kV/mm, as indicated in the plot. (b) Variation of polarization represented as a function of the applied field using the poling time as parameter. (c) Derivative of polarization with respect to the applied field. (d) Normalized derivative of polarization with respect to the applied field *vs.* the normalized applied field.

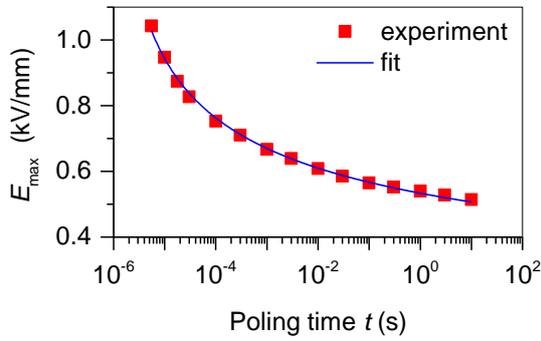

Figure 3. The maximum position $E_{max}$ versus poling time $t$. Symbols correspond to the experimental data, whereas the solid line represents the fit by the expression $E_{max}(t) = E_0 / [\ln(t/\tau_0)]^{1/\alpha}$, with $E_{a\_}$ = 1.28±0.010 kV/mm, $\tau_0$ = (8.00±0.03) 10⁻⁷ s, $\alpha$ = 3.010±0.035.

The dependence of $E_{max}(t)$ in Figure 3 is very well fitted by the expression (18) derived from the Merz law [13], $E_{max}(t) = E_0 / [\ln(t/\tau_0)]^{1/\alpha}$, using the parameters $E_0 = 1.28$ kV/mm, $\tau_0 = 8 \times 10^{-7}$ s and $\alpha = 3.01$ with an overall error of about 0.2%. By implementing this dependence and the master curve of Figure 2(d) in the numerical form, theoretical polarization response according to Eq. (2) is evaluated for the respective applied field values and represented in Figure 2(a) by solid lines. Theoretical curves mostly describe the experiment with high accuracy except for some deviations for the highest field values and shortest measured times.

We note that, for ferroelectric ceramics of tetragonal and rhombohedral phase symmetries, the expressions for the total polarization, Eqs. (6,16), and the master curve $\Phi(u)$, Eq. (19), are different (see in Appendices A and B), but the principal fact of scaling properties and the validity of Eqs. (1,2) remains. This is confirmed by experimental observations in Refs. [29,46].

## 4. Conclusions

Dynamic scaling properties of polarization response in multiaxial polycrystalline ferroelectrics, exhibiting parallel and sequential switching events, are predicted by the combined MSM-IFM model and exemplarily experimentally proven on the orthorhombic KNN-based ceramic. The analysis allows insight into microscopic switching mechanisms, represented by activation fields of individual switching events, on the one hand, and into macroscopic statistical properties of the polycrystalline media, represented by the weighted statistical distribution of electric field in them, on the other hand. This knowledge substantially improves the description of the temporal polarization response. Thus the universal properties of the polarization reversal, observed before in various polycrystalline ferroelectrics undergoing one-step polarization switching events, are extended to a wider class of multiaxial ferroelectrics undergoing multistep polarization processes. This approach can be readily extended to further crystalline classes of ferroelectrics, undergoing thermally activated field-driven polarization switching, like hexagonal and monoclinic materials. Furthermore, the insight in multistep polarization switching kinetics is crucial for future ultrafast switching applications and, in particular, for multilevel digital information storage.

**Acknowledgements**

This work was supported by the Deutsche Forschungsgemeinschaft (DFG) Grants Nos. 405631895 (KO 5100/1-1) and GR 479/2. K.W. acknowledges the support of the National Nature Science Foundation of China (Grant Nos. 51822206 and 52032005).

**Data Availability**

Data available on request from the authors.

**Appendix A. Scaling properties of polycrystalline tetragonal ferroelectrics**

The multistep polarization reversal in tetragonal ferroelectrics was studied in Ref. [43] where possible one-step 180°- and two-step 90°-90°-switching events were accounted, resulting in the total time-dependent polarization variation

$$\Delta P(t) = \Delta P_{180}(t) + \Delta P_{90-90}(t). \tag{A1}$$

Using the probabilities of different switching steps derived previously [43] and applying to them the step-function approximation, the formula (A1) for single-crystalline tetragonal ferroelectrics is transformed to

$$\Delta P(t) \cong \eta P_s \left[ \vartheta(t-\tau_1) + \vartheta(t-\tau_1-\tau_2) \right] + (1-\eta) 2 P_s \vartheta(t-\tau_3) \tag{A2}$$

where, in notations of Ref. [43], $\tau_1, \tau_2$ and $\tau_3$ are the switching times, of the first 90°, the second 90° and the 180° switching steps, respectively, and $\eta$ is the fraction of two-step 90°-90° switching events. Considering the fact that $\tau_1 \ll \tau_2, \tau_3$ in the second theta-function an approximation $t-\tau_1-\tau_2 \cong t-\tau_2$ can be adopted.

In order to describe the polarization response of polycrystalline ferroelectrics in the IFM approach, the formula (A2) should be averaged over the statistical distribution of local electric films. To this end, the step-functions on the time scale should be first converted into step-functions on the field scale as follows:

$$\Delta P(t) \cong \eta P_s \left[ \vartheta\left(E-E_{th}^{(1)}(t)\right) + \vartheta\left(E-E_{th}^{(2)}(t)\right) \right] + (1-\eta) 2 P_s \vartheta\left(E-E_{th}^{(3)}(t)\right). \tag{A3}$$

where the introduced threshold fields $E_{th}^{(n)}(t)$ are the solutions of respective equations $\tau_n(E) = t$.

After averaging of Eq. (A3) using the weighted statistical distribution function, Eq. (3), and an applied field $E_a$, one obtains

$$\frac{\Delta P(E_a, t)}{\Delta P_{max}} \cong \frac{\eta}{2} \int_{E_{th}^{(1)}(t)/E_a}^{\infty} ds\, f(s) + \frac{\eta}{2} \int_{E_{th}^{(2)}(t)/E_a}^{\infty} ds\, f(s) + (1-\eta) \int_{E_{th}^{(3)}(t)/E_a}^{\infty} ds\, f(s). \tag{A4}$$

Then the normalized derivative reads

$$\frac{E_a}{\Delta P_{max}} \frac{\partial \Delta P(E_a,t)}{\partial E_a} \cong \frac{\eta}{2} \frac{E_{th}^{(1)}}{E_a} f\left(\frac{E_{th}^{(1)}}{E_a}\right) + \frac{\eta}{2} \frac{E_{th}^{(2)}}{E_a} f\left(\frac{E_{th}^{(2)}}{E_a}\right) + (1-\eta) \frac{E_{th}^{(3)}}{E_a} f\left(\frac{E_{th}^{(3)}}{E_a}\right). \tag{A5}$$

Assuming that the Merz law, Eq. (18), applies for all switching times $\tau_n$, all terms in the right-hand side of Eq. (A5) turn out to depend on the combined variable $E_a \left[ \ln(t/\tau_0) \right]^{1/\alpha}$. It applies then also to the left-hand side of Eq. (A5) and its maximum $E_{\max}(t)$. This provides the relation $E_{\max} = \gamma_1 E_{th}^{(1)} = \gamma_2 E_{th}^{(2)} = \gamma_3 E_{th}^{(3)}$ with unknown constants $\gamma_n$ and finally results in the scaling relation, Eq. (3), with the master curve

$$\Phi(s) = \frac{\eta}{2} \frac{1}{\gamma_1 s} f\left(\frac{1}{\gamma_1 s}\right) + \frac{\eta}{2} \frac{1}{\gamma_2 s} f\left(\frac{1}{\gamma_2 s}\right) + (1-\eta) \frac{1}{\gamma_3 s} f\left(\frac{1}{\gamma_3 s}\right). \tag{A6}$$

For establishing the 90°-process fraction $\eta$ and constants $\gamma_n$ additional information is required, such as measurements of the time-dependent strain response [42-46]. We note that, in the experimentally studied case of a tetragonal PZT ceramic [43], the relation $\tau_2 = \tau_3$, implying also $E_{th}^{(2)} = E_{th}^{(3)}$ and $\gamma_2 = \gamma_3$, was observed which further simplifies the expression (A6). We again emphasize that, for describing the polarization response using Eq. (2), the knowledge of process fractions and respective constants $\gamma_n$ is not required, as soon as the scaling relation (1) is experimentally established as it was done in Ref. [29].

**Appendix B. Scaling properties of polycrystalline rhombohedral ferroelectrics**

The multistep polarization reversal in rhombohedral ferroelectrics was studied in Ref. [44] where possible parallel one-step 180°- and sequential two-step 109°-71°-, 71°-109°- and three-step 71°-71°-71°-switching processes were accounted, resulting in the total time-dependent polarization variation

$$\Delta P(t) = \Delta P_{180}(t) + \Delta P_{109-71}(t) + \Delta P_{71-109}(t) + \Delta P_{71-71-71}(t). \tag{B1}$$

Using the probabilities of different switching steps derived previously [44] and applying to them the step-function approximation, the contributions to the formula (B1) for single-crystalline rhombohedral ferroelectrics can be represented as

$$\Delta P_{180}(t) \cong \eta_3 2 P_s \vartheta(t-\tau_3)$$
$$\Delta P_{109-71}(t) \cong \eta_2 P_s \left[ \frac{4}{3}\vartheta(t-\tau_2) + \frac{2}{3}\vartheta(t-\tau_2-\tau_{21}) \right]$$
$$\Delta P_{71-109}(t) \cong \eta_{12} P_s \left[ \frac{2}{3}\vartheta(t-\tau_1) + \frac{4}{3}\vartheta(t-\tau_1-\tau_{12}) \right]$$
$$\Delta P_{71-71-71}(t) \cong \eta_{11} P_s \frac{2}{3}\left[ \vartheta(t-\tau_1) + \vartheta(t-\tau_1-\tau_{11}) + \vartheta(t-\tau_1-\tau_{11}-\tau_{111}) \right]$$
(B2)

where, in notations of Ref. [44], $\tau_1$, $\tau_{11}$ and $\tau_{111}$ are the switching times of the first 71°, the second 71° and the third 71° switching steps, respectively, $\tau_2$ and $\tau_{21}$ are the switching times of the first 109° and the following 71° switching steps, respectively, $\tau_{12}$ is the switching time of the second 109° switching step after the first 71° one and $\tau_3$ is the switching time of the one-step 180° switching process. The parameters $\eta_{11}, \eta_{12}, \eta_2$ and $\eta_3$ present the fractions of the three-step 71°-71°-71°, the two-step 71°-109°, the two-step 109°-71° and the one-step 180° switching paths, respectively. The sequential switching times typically differ by orders of the magnitude. In the case of the studied rhombohedral PZT ceramic [44], the following relations were established: $\tau_1 \ll \tau_{12}$, $\tau_2 \ll \tau_{21}$ and $\tau_1 \ll \tau_{11} \ll \tau_{111}$. This allows simplification of some arguments in Eq. (B2): $t-\tau_1-\tau_{12} \cong t-\tau_{12}$, $t-\tau_2-\tau_{21} \cong t-\tau_{21}$, $t-\tau_1-\tau_{11} \cong t-\tau_{11}$, $t-\tau_1-\tau_{11}-\tau_{111} \cong t-\tau_{111}$.

In order to describe the polarization response of polycrystalline ferroelectrics in the IFM approach, the formulas (B2) should be averaged over the statistical distribution of local electric films. To this end, the step-functions on the time scale should be first converted into step-functions on the field scale. This leads to the forms

$$\Delta P_{180}(t) \cong \eta_3 2 P_s \vartheta(E-E_{th}^{(3)}(t))$$
$$\Delta P_{109-71}(t) \cong \eta_2 P_s \left[ \frac{4}{3}\vartheta(E-E_{th}^{(2)}(t)) + \frac{2}{3}\vartheta(E-E_{th}^{(21)}(t)) \right]$$
$$\Delta P_{71-109}(t) \cong \eta_{12} P_s \left[ \frac{2}{3}\vartheta(E-E_{th}^{(1)}(t)) + \frac{4}{3}\vartheta(E-E_{th}^{(12)}(t)) \right]$$
$$\Delta P_{71-71-71}(t) \cong \eta_{11} P_s \frac{2}{3}\left[ \vartheta(E-E_{th}^{(1)}(t)) + \vartheta(E-E_{th}^{(11)}(t)) + \vartheta(E-E_{th}^{(111)}(t)) \right]$$
(B3)

where the introduced threshold fields $E_{th}^{(n)}(t)$ are the solutions of respective equations $\tau_n(E) = t$.

After averaging of Eq. (B3) using the weighted statistical distribution function, Eq. (3), and an applied field $E_a$, one obtains

$$\frac{\Delta P(E_a,t)}{\Delta P_{\max}} \cong \frac{1}{3}(\eta_{11}+\eta_{12}) \int_{E_{th}^{(1)}(t)/E_a}^{\infty} ds\, f(s) + \frac{1}{3}\eta_{11} \int_{E_{th}^{(11)}(t)/E_a}^{\infty} ds\, f(s) + \frac{1}{3}\eta_{11} \int_{E_{th}^{(111)}(t)/E_a}^{\infty} ds\, f(s)$$
$$+ \frac{2}{3}\eta_{12} \int_{E_{th}^{(12)}(t)/E_a}^{\infty} ds\, f(s) + \frac{2}{3}\eta_{2} \int_{E_{th}^{(2)}(t)/E_a}^{\infty} ds\, f(s) + \frac{1}{3}\eta_{2} \int_{E_{th}^{(21)}(t)/E_a}^{\infty} ds\, f(s) + \eta_{3} \int_{E_{th}^{(3)}(t)/E_a}^{\infty} ds\, f(s). \tag{B4}$$

Then the normalized derivative reads

$$\frac{E_a}{\Delta P_{\max}} \frac{\partial \Delta P(E_a,t)}{\partial E_a} \cong \frac{1}{3}(\eta_{11}+\eta_{12})\frac{E_{th}^{(1)}}{E_a} f\!\left(\frac{E_{th}^{(1)}}{E_a}\right) + \frac{1}{3}\eta_{11}\frac{E_{th}^{(11)}}{E_a} f\!\left(\frac{E_{th}^{(11)}}{E_a}\right) + \frac{1}{3}\eta_{11}\frac{E_{th}^{(111)}}{E_a} f\!\left(\frac{E_{th}^{(111)}}{E_a}\right)$$
$$\frac{2}{3}\eta_{12}\frac{E_{th}^{(12)}}{E_a} f\!\left(\frac{E_{th}^{(12)}}{E_a}\right) + \frac{2}{3}\eta_{2}\frac{E_{th}^{(2)}}{E_a} f\!\left(\frac{E_{th}^{(2)}}{E_a}\right) + \frac{1}{3}\eta_{2}\frac{E_{th}^{(21)}}{E_a} f\!\left(\frac{E_{th}^{(21)}}{E_a}\right) + \eta_{3}\frac{E_{th}^{(3)}}{E_a} f\!\left(\frac{E_{th}^{(3)}}{E_a}\right).$$
$$\tag{B5}$$

Assuming that the Merz law, Eq. (18), applies for all switching times $\tau_n$, all terms in the right-hand side of Eq. (B5) turn out to depend on the combined variable $E_a\left[\ln(t/\tau_0)\right]^{1/\alpha}$. It applies then also to the left-hand side of Eq. (B5) and its maximum $E_{\max}(t)$. This provides the relation $E_{\max} = \gamma_1 E_{th}^{(1)} = \gamma_{11} E_{th}^{(11)} = \gamma_{111} E_{th}^{(111)} = \gamma_{12} E_{th}^{(12)} = \gamma_2 E_{th}^{(2)} = \gamma_{21} E_{th}^{(21)} = \gamma_3 E_{th}^{(3)}$ with unknown constants $\gamma_n$ and finally results in the scaling relation, Eq. (3), with the master curve

$$\Phi(s) = \frac{1}{3}(\eta_{11}+\eta_{12})\frac{1}{\gamma_1 s} f\!\left(\frac{1}{\gamma_1 s}\right) + \frac{1}{3}\eta_{11}\frac{1}{\gamma_{11} s} f\!\left(\frac{1}{\gamma_{11} s}\right) + \frac{1}{3}\eta_{11}\frac{1}{\gamma_{111} s} f\!\left(\frac{1}{\gamma_{111} s}\right)$$
$$+\frac{2}{3}\eta_{12}\frac{1}{\gamma_{12} s} f\!\left(\frac{1}{\gamma_{12} s}\right) + \frac{2}{3}\eta_{2}\frac{1}{\gamma_{2} s} f\!\left(\frac{1}{\gamma_{2} s}\right) + \frac{1}{3}\eta_{2}\frac{1}{\gamma_{21} s} f\!\left(\frac{1}{\gamma_{21} s}\right) + \eta_{3}\frac{1}{\gamma_{3} s} f\!\left(\frac{1}{\gamma_{3} s}\right). \tag{B6}$$

For determining the process fractions $\eta_n$ and constants $\gamma_n$ additional information is required, such as measurements of the time-dependent strain response [42-46]. We note that, in the experimentally studied case of a rhombohedral PZT ceramic [44], the relations $\tau_{111} = \tau_{21} \cong \tau_{12} = \tau_3$, implying also $E_{th}^{(111)} = E_{th}^{(21)} \cong E_{th}^{(12)} = E_{th}^{(3)}$ and $\gamma_{111} = \gamma_{21} = \gamma_{12} = \gamma_3$, were established which further simplifies the expression (B6). We again emphasize that, for describing the polarization response using Eq. (2), the knowledge of process fractions $\eta_n$ and respective constants $\gamma_n$ is not required, as soon as the scaling relation (1) is experimentally established as it was done in Ref. [29].

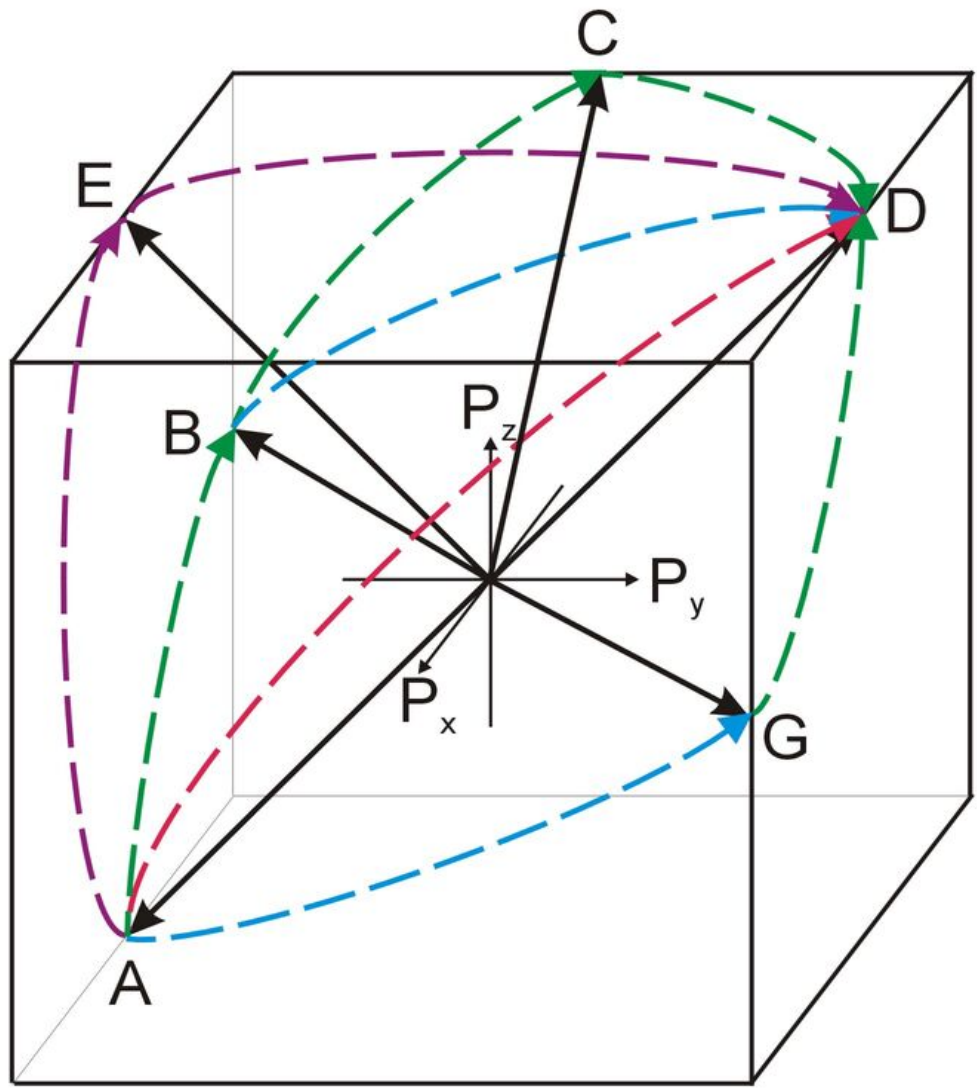

(a)

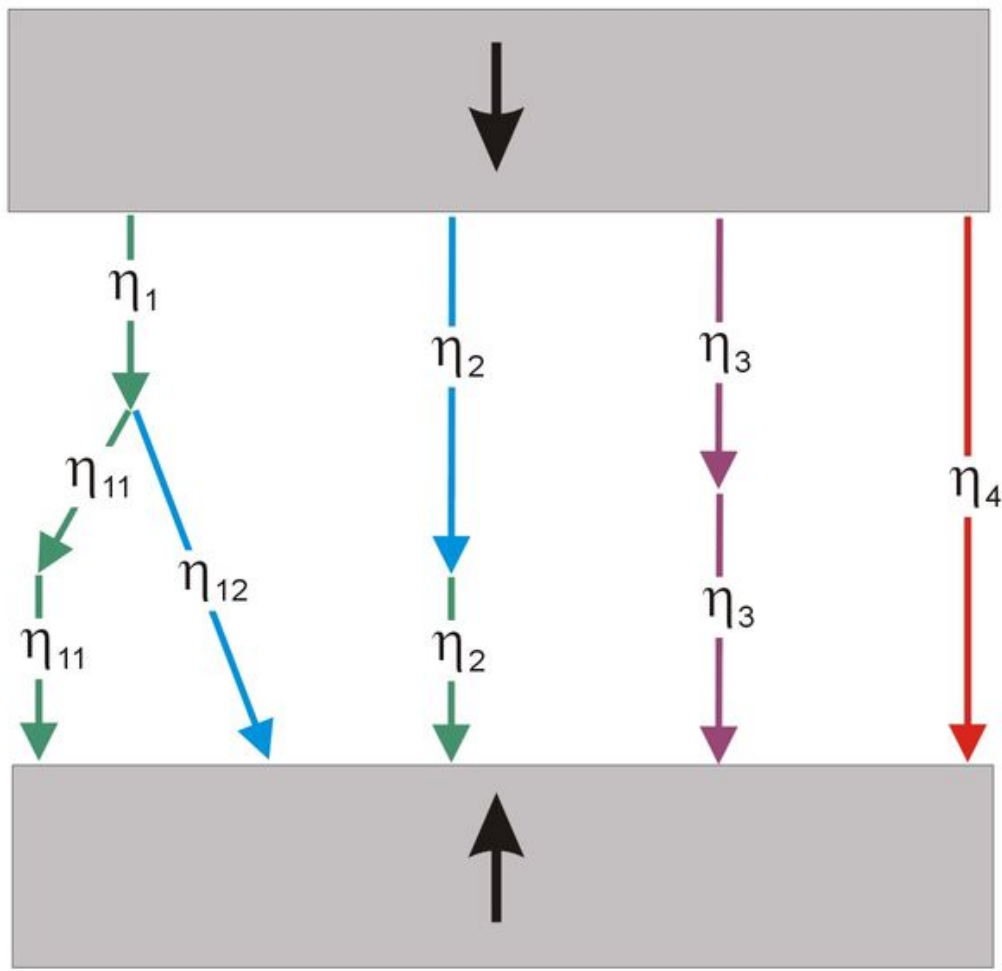

(b)

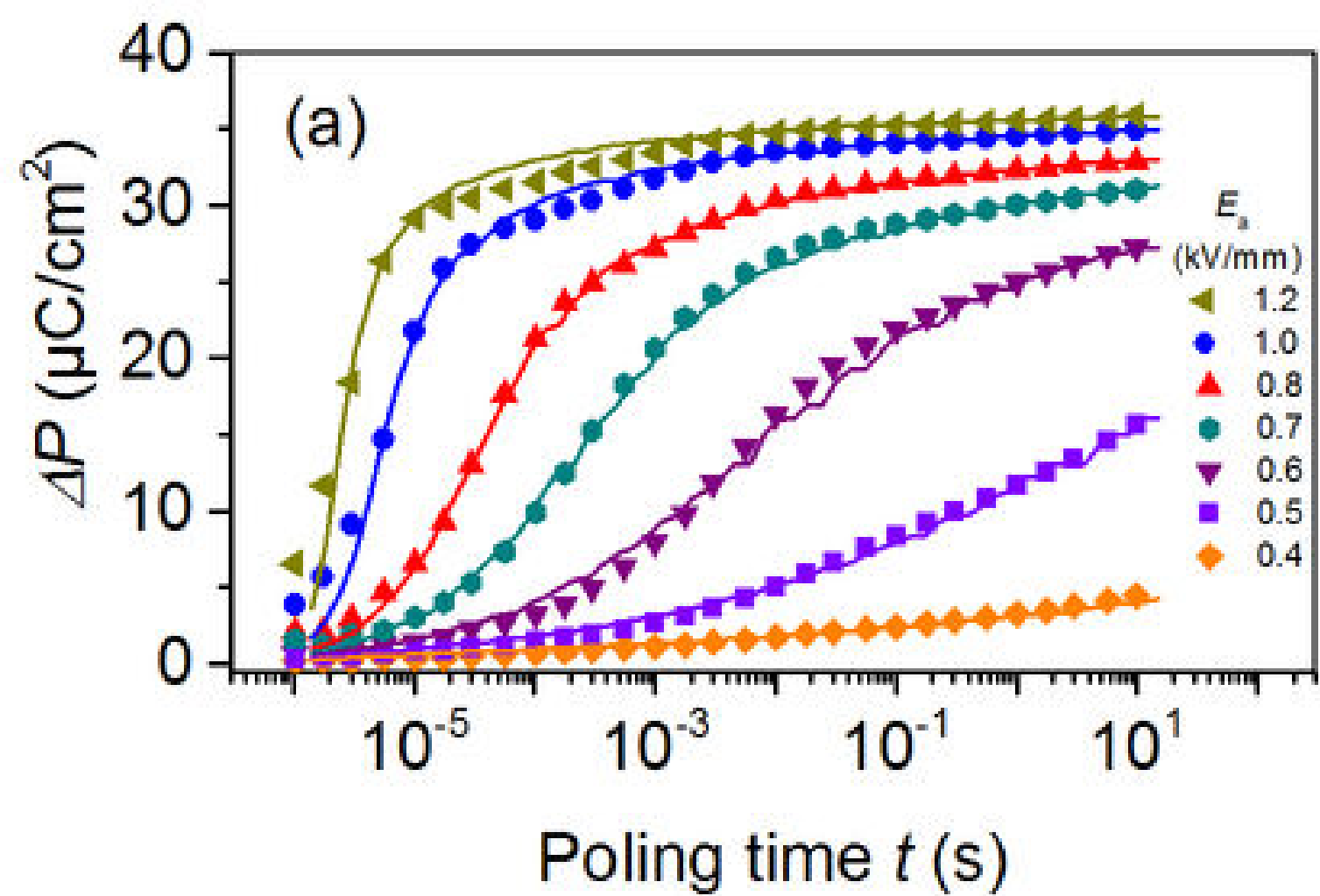

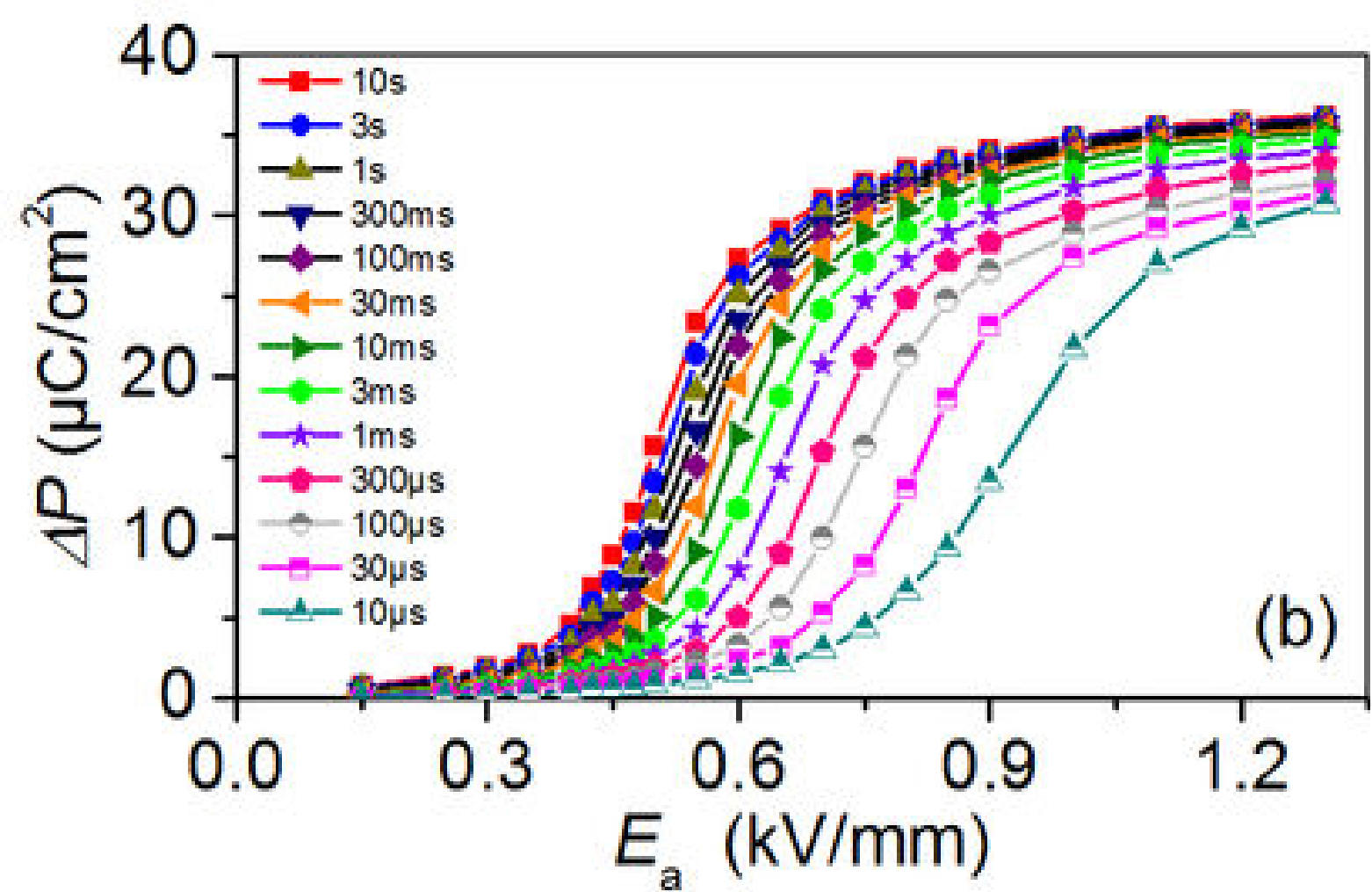

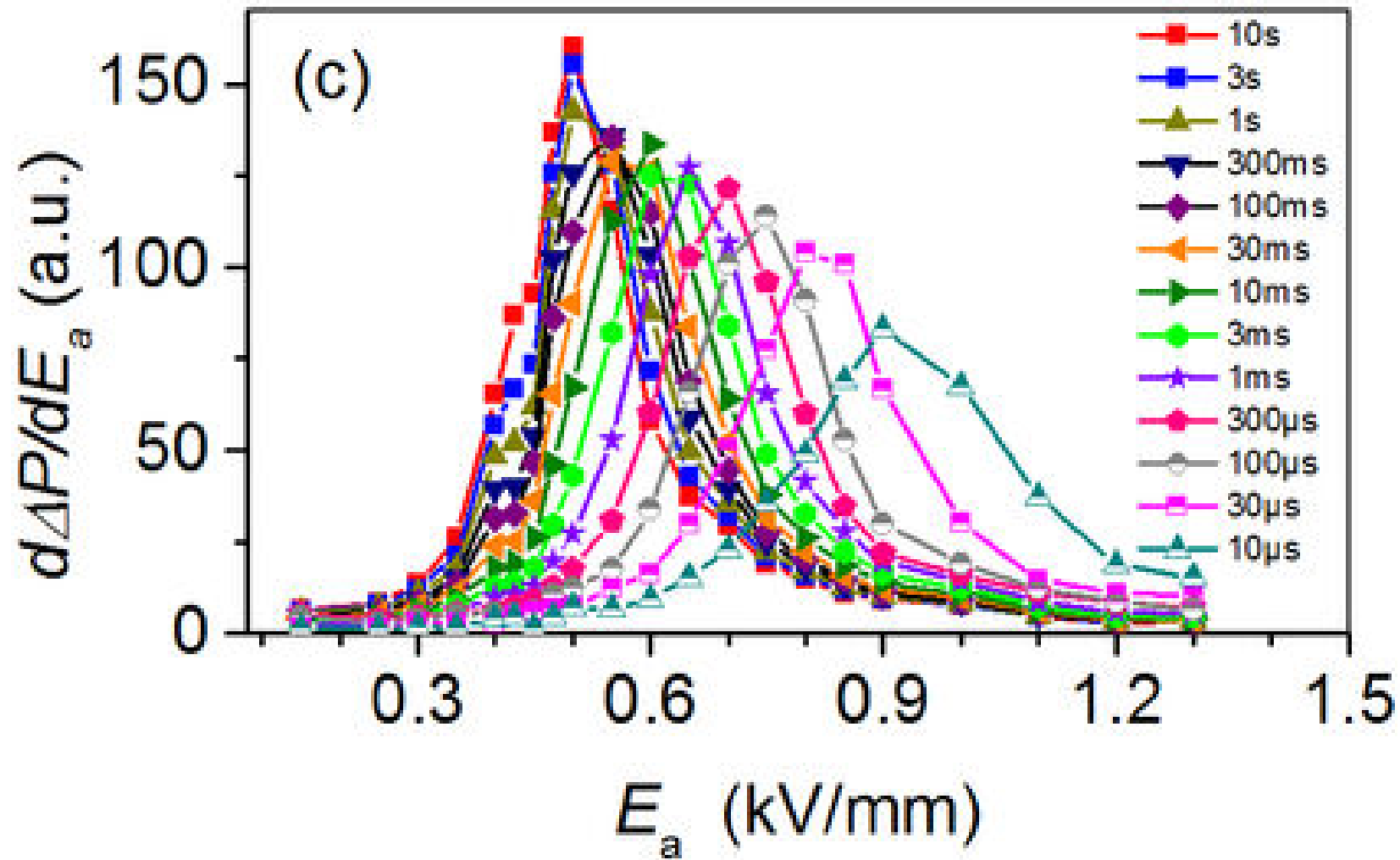

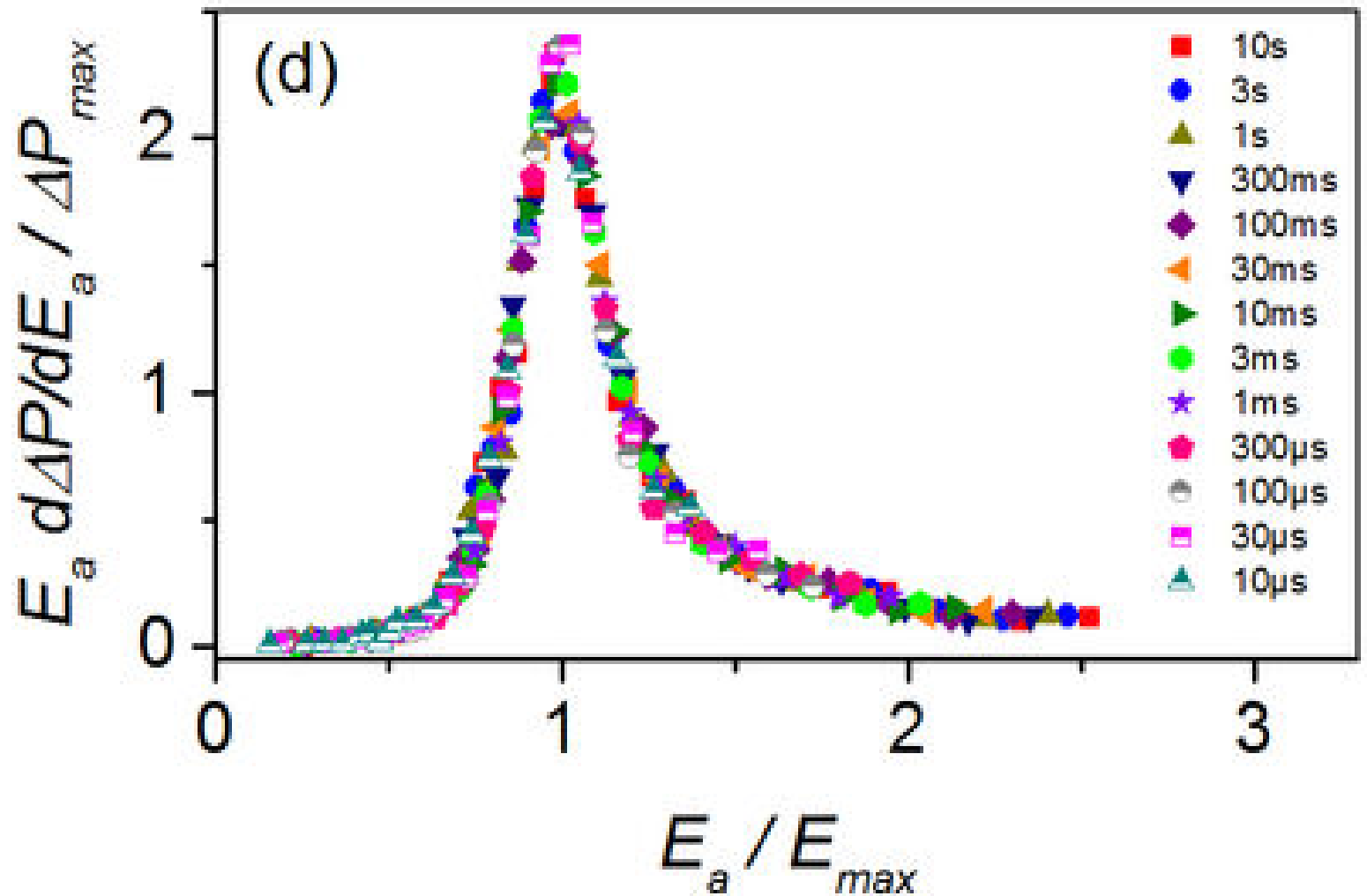

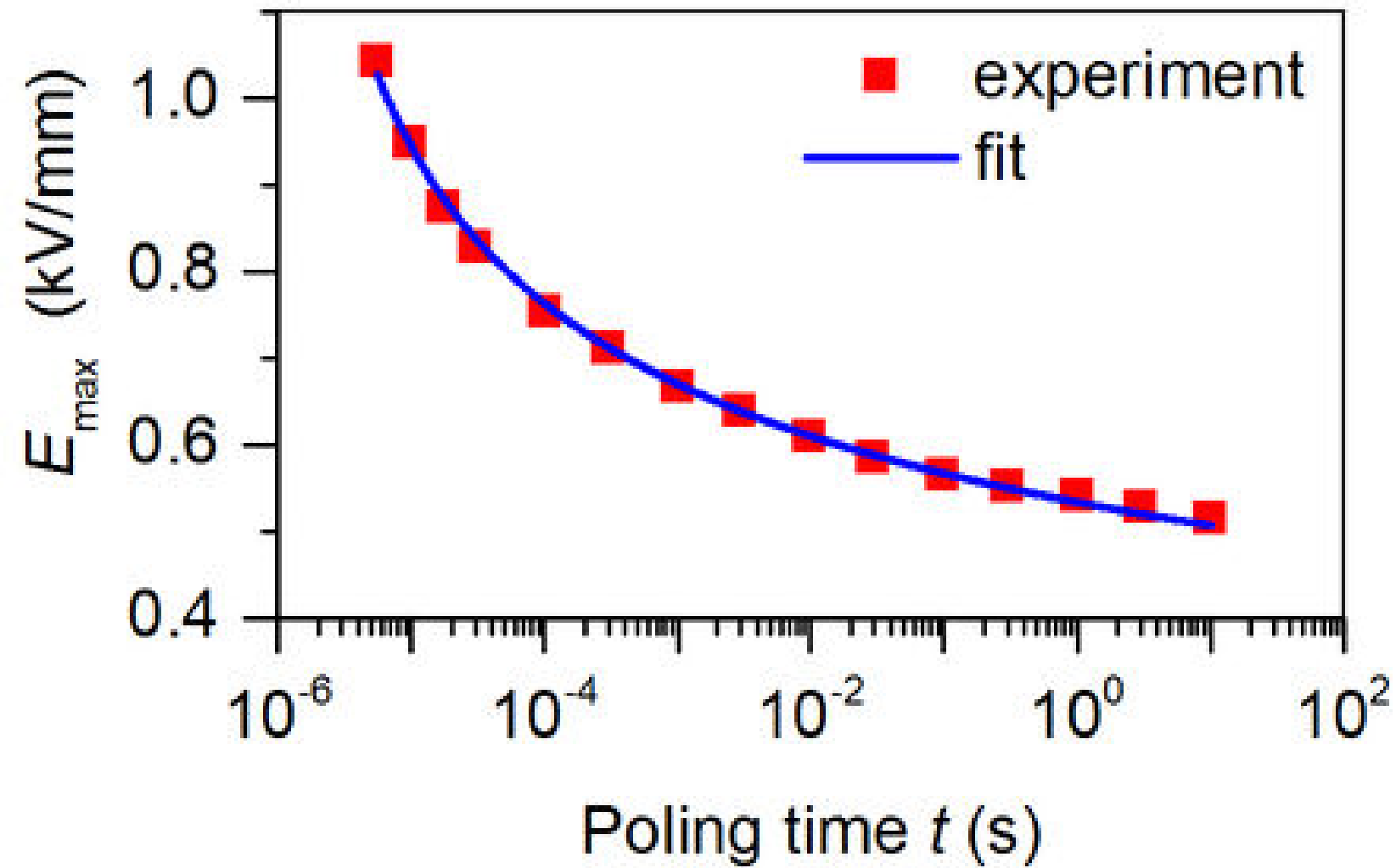